\documentclass[
    ,final            
  ]
  {aipproc}

\def\stau{\widetilde{\tau}}
\def\snu{\widetilde{\nu}}
\def\mstau{m_{\stau}}
\def\gravitino{\widetilde{G}}
\def\mgravitino{m_{\gravitino}}

\def\ve{\varepsilon}

\newcommand{\gtrsim}{ \mathop{}_{\textstyle \sim}^{\textstyle >} }
\newcommand{\lesssim}{ \mathop{}_{\textstyle \sim}^{\textstyle <} }

\def\mt{\widetilde{m}_1}

\layoutstyle{6x9}

\begin{document}

\title{Leptogenesis and Gravitino Dark Matter}

\classification{98.80.Cq, 95.35.+d, 12.60.Jv, 95.30.Cq}
\keywords      {Baryogenesis, Dark Matter, Big Bang Nucleosynthesis}

\author{Wilfried Buchm\"uller}
{address={Deutsches Elektronen-Synchroton DESY, 22607 Hamburg, Germany}}

\begin{abstract}
We study implications of thermal leptogenesis for the superparticle mass
spectrum. A consistent picture is obtained if the lightest superparticle
is the gravitino, which can be the dominant component of cold dark
matter. In the case of a long-lived charged scalar lepton as next-to-lightest
superparticle, supergravity can be tested at the next generation of
colliders, LHC and ILC. 
\end{abstract}

\maketitle

\section{Introduction}

Leptogenesis \cite{fy86} is an attractive theory for the origin of matter.
Heavy Majorana neutrinos, the seesaw partners of the known light neutrinos,
naturally generate a B-L asymmetry in CP violating out-of-equilibrium decays
which, via sphaleron processes, is partially converted into a baryon asymmetry.
Thermal leptogenesis, where the generated baryon asymmetry is closely
related to the light neutrino masses, requires a rather high temperature 
in the early universe.

In supersymmetric extensions of the standard model, there is a potential
clash between the required baryogenesis temperature and the thermal 
overproduction of gravitinos whose late decays alter the standard predictions
of primordial nucleosynthesis (BBN). This leads to strong constraints on the
allowed superparticle mass spectrum and, in particular, the allowed lightest
superparticle (LSP) \cite{bpy05}.

A consistent picture can be obtained in models where the gravitino is the
LSP. In a certain mass range for the gravitino and the next-to-lightest
superparticle (NLSP) one can avoid the BBN constraints and gravitinos can
be the dominant component of dark matter. The NLSP is then quasi-stable,
which leads to very interesting effects at colliders \cite{fxx06}.

After some comments on the current status of leptogenesis, we discuss in
the following the BBN constraints on gravitinos and scalar lepton NLSPs,
the perspective for explaining dark matter in terms of gravitinos and
some collider signatures.

\section{STATUS OF LEPTOGENESIS}

During the past years the theory of leptogenesis has reached a remarkable
quantitative state, especially for the simplest scenario where the generation
of the baryon asymmetry is dominated by the decays of the lightest of the
heavy Majorana neutrinos (`$N_1$-dominance'). In this case the ratio of
baryon density to photon density is determined just by the CP asymmetry 
$\ve_1$ in $N_1$ decays, which depends on neutrino masses and mixings, and 
the efficiency factor $\kappa_f$, which accounts for the dynamics of the 
non-equilibrium processes in the plasma of the early universe,
\begin{equation}
\eta_B = {n_B\over n_\gamma} \simeq 0.01\, \ve_1\, \kappa_f\;.
\end{equation}
In Fig.~1 the efficiency factor $\kappa_f$ is shown as function of the 
`effective neutrino mass' $\mt$ which varies between the smallest and largest 
light neutrino mass, $m_1 \leq \mt \lesssim m_3$. For values below the
`equilibrium neutrino mass', $\mt < m_* \simeq 10^{-3}\ {\rm eV}$, the
efficiency factor, and therefore the predicted baryon asymmetry has a
large uncertainty. Depending on the initial abundance of the heavy neutrinos
$N_1$ and theoretical uncertainties, it varies over several orders of
magnitude. On the contrary, for $\mt > m_*$, which corresponds to the
solar and atmospheric neutrino masses, the value of $\kappa_f$ is rather
accurately determined. In order to avoid too strong lepton-number erasing
`washout processes', the neutrino masses have to obey also the upper bound
$m_i < 0.1\ {\rm eV}$. Altogether one then obtains the light neutrino mass
window for successful leptogenesis \cite{bdp03}
\begin{equation}
10^{-3}\ {\rm eV} < m_i < 0.1\ {\rm eV}\;,
\end{equation}
which will soon be probed by laboratory experiments and cosmological 
observations. The quantitative analysis of leptogenesis also yields a
lower bound on the baryogenesis temperature \cite{di02,bdp04},
\begin{equation} 
T_B \sim M_1 \gtrsim 10^9\ {\rm GeV}\ ,
\end{equation}
which is very important for the following discussion on gravitino dark matter.

\begin{figure}
\begin{tabular}{c}
  \includegraphics[height=.35\textheight]{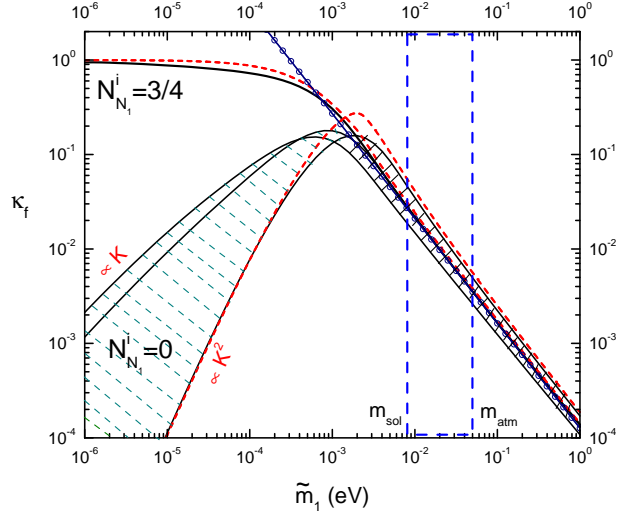} 
  \caption{The efficiency factor $\kappa_f$ as function of the effective
neutrino mass $\mt$. From \cite{bdp04}.}
\end{tabular}
  \label{prime}
\end{figure}

There are several possibilities to avoid the constraints of `standard thermal
leptogenesis' on light neutrino masses and the reheating temperature: the
minimal seesaw mechanism can be modified by adding SU(2) triplet fields,
one can make use of the enhanced CP asymmetry in the case of degenerate
heavy neutrino masses (`resonant leptogenesis') or one can use non-thermal
processes to generate the initial heavy neutrino abundance \cite{bpy05,ham04}.

An important recent development concerns the detailed study of flavour
effects in standard thermal leptogenesis, which can strongly affect
the light neutrino mass bound \cite{abx06}. The constraints on the 
superparticle mass spectrum follow from the lower bound on the reheating
temperature, which is only mildly relaxed by flavour effects.

\begin{figure}
\begin{tabular}{c}
  \includegraphics[height=.40\textheight]{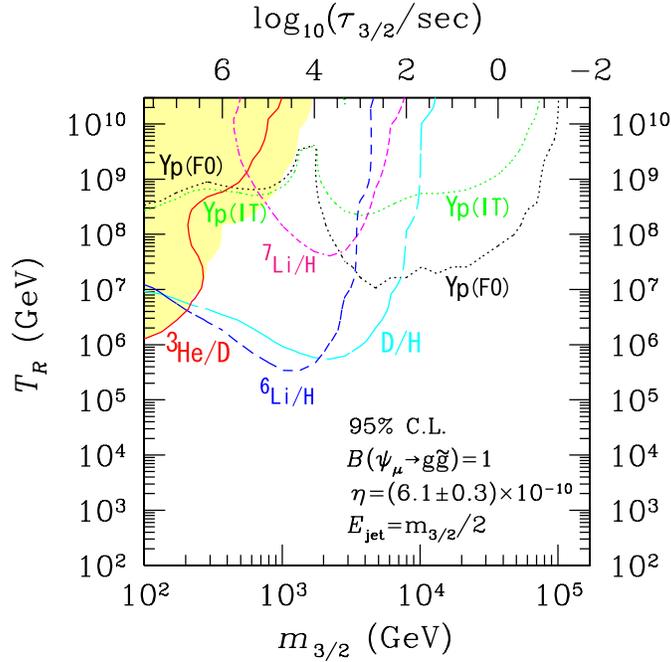} 
  \caption{Upper bounds on the reheating temperature as function of the
gravitino mass. From \cite{kkm05}.}
\end{tabular}
\end{figure}

\section{Constraints from nucleosynthesis (BBN)}

In a supersymmetric plasma at high temperature gravitinos are thermally
produced, mostly by QCD processes. Their number density $n_{3/2}$ increases 
linearly with the reheating temperature,
\begin{equation}
{n_{3/2}\over n_\gamma} \propto {\alpha_3\over M_{\rm p}^2}\,T_R\;,
\end{equation}
where $M_{\rm p}$ and $\alpha_3$ are the Planck mass and the QCD fine
structure constant, respectively. The late decay of the gravitinos 
alters the successful BBN prediction, which implies upper bounds for the
reheating temperature $T_R$. The most stringent one \cite{kkm05} 
shown in Fig.~2 yields
\begin{equation}
T_R < {\cal O}(1)\times 10^5\ {\rm GeV}\ ,
\end{equation}
which is clearly incompatible with the lower bound from thermal leptogenesis!

\begin{figure}
\begin{tabular}{c}
  \includegraphics[height=.25\textheight]{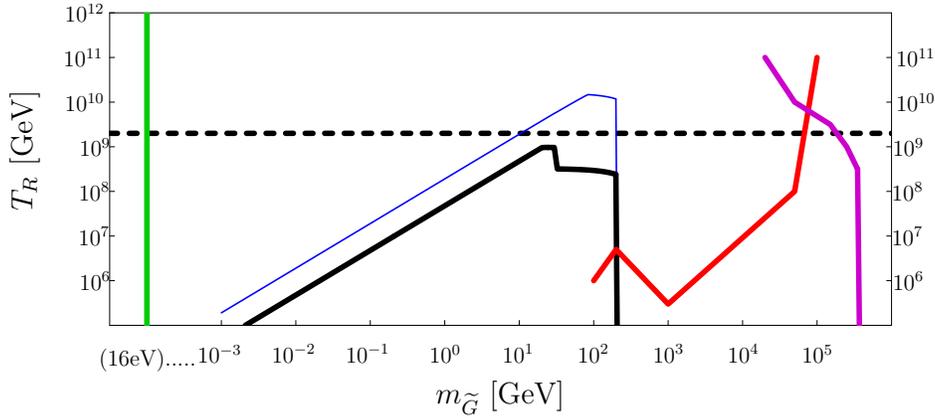} 
  \caption{Constraints on the reheating temperature from BBN and 
$\Omega_{\rm CDM}$ as function of the gravitino mass. The dashed line
is the lower bound on $T_R$ from leptogenesis. Allowed regions of the gravitino
mass are below $16\ {\rm eV}$, around $50\ {\rm GeV}$ and around 
$10^5\ {\rm GeV}$. The blue (black) line corresponds to 
$m_{\widetilde{g}} = 500\ {\rm GeV}$ ($1\ {\rm TeV}$). The typical gravitino 
mass ${\cal O}(1\ {\rm TeV})$ in supergravity models is inconsistent with 
leptogenesis.}
\end{tabular}
  \label{sqgl}
\end{figure}

The conflict between the upper bound from BBN and the lower bound from
leptogenesis on the reheating temperature can be avoided if the gravitino
is the LSP \cite{bbp98}. In this case the BBN bounds apply to the NLSP which
is quasi-stable. The gravitino production is now enhanced,
\begin{equation}
{n_{3/2}\over n_\gamma} \propto {\alpha_3\over M_{\rm p}^2}\,
\left({m_{\widetilde{g}}\over m_{3/2}}\right)^2\,T_R\ ,
\end{equation}
where $m_{\widetilde{g}}$ and $m_{3/2}$ are gluino and gravitino mass,
respectively. The requirement that the gravitino density does not exceed
the observed $\Omega_{\rm DM}$ yields a strong constraint on the reheating
temperature for light gravitinos \cite{mmy93}. Cosmological observations 
constrain
very light gravitinos to have masses below $16\ {\rm eV}$ \cite{vie05}.
BBN and $\Omega_{\rm DM}$ constraints require unstable gravitinos to be
extremely heavy \cite{ibe05}, $m_{3/2} \sim 10^5\ {\rm GeV}$. The allowed
regions for the gravitino mass are summarized in Fig.~3. The BBN constraints
for the NLSP have recently been studies by several groups \cite{sev04,ste06}.
A neutralino NLSP is excluded, the particularly interesting case of a 
$\stau$ NLSP is strongly constrained, and a $\snu$ NLSP is essentially
unconstrained. Recently it has been argued that quasi-stable charged particles 
alter BBN  via the formation of bound states \cite{pos06,kt06}. This
could lead to further very restrictive constraints on the allowed
NLSP abundance.

\begin{figure}
\begin{tabular}{c}
  \includegraphics[height=.35\textheight]{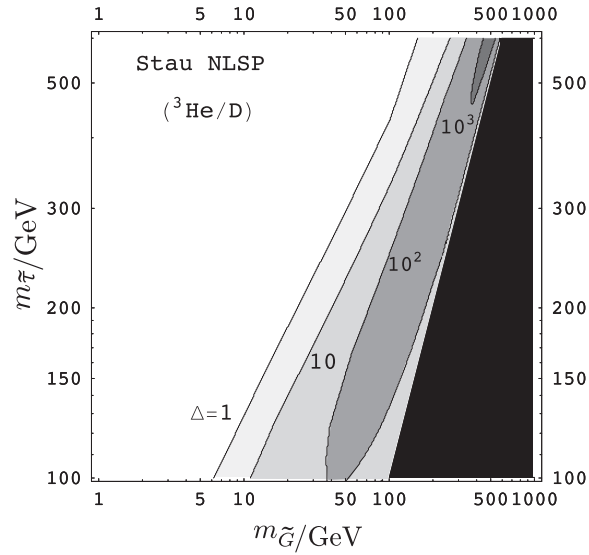} 
  \caption{The BBN constraint (${}^{3}$He/D bound) on the parameter space 
  $( \mgravitino, \mstau)$ with late-time entropy production.
 The regions excluded by the ${}^{3}$He/D bound
 are shaded from light to dark gray 
 for a dilution factor $\Delta=1$, 10, $10^{2}$, $10^{3}$.
 In the black shaded region, the gravitino is not the LSP.
 The effects of hadronic decays have been neglected. From \cite{ibe06}.} 
\end{tabular}
  \label{stop}
\end{figure}

The BBN constraints can be weakened by late-time entropy
production, i.e., after NLSP decoupling and before BBN. The `relic'
thermal density of $\stau$'s, $Y_{\stau}^{\mathrm{thermal}} = 
n_{\stau}/{s}$, is then diluted, 
\begin{equation}
Y_{\stau} = \frac{1}{\Delta} Y_{\stau}^{\mathrm{thermal}}\ .
\end{equation}
This considerably enlarges the parameter region with relatively large
gravitino mass, $\mgravitino \geq {\cal O}(0.1)\,m_{\stau}$, which is 
of particular interest for testing supergravity at colliders. The allowed
regions in the $\mgravitino$-$m_{\stau}$-plane are shown in Fig.~4 for 
different values of $\Delta$.

\section{GRAVITINO DARK MATTER}

So far we have discussed the constraints which the lower bound on the
reheating temperature from leptogenesis imposes on the superparticle mass
spectrum, and we have argued that the gravitino LSP with an appropriate
NLSP represents a consistent scenario. This then leads to the question
whether one can understand the observed amount of cold dark matter,
$\Omega_{\rm DM}h^2 = \rho_{\rm DM}h^2/\rho_c \simeq 0.11$,
in terms of gravitinos, i.e., $\Omega_{\rm DM} \simeq \Omega_{3/2}$.

One possible explanation is the SuperWIMP mechanism \cite{frt03}
where gravitinos are mainly produced in WIMP decays. The gravitino mass
density is then determined by the NLSP density,
\begin{equation}
\Omega_{3/2} = {m_{3/2}\over m_{\rm NLSP}} \Omega_{\rm NLSP}\;,
\end{equation}
which is independent of the reheating temperature $T_R$! The BBN constraints
require, however, a rather large $\stau$ mass, $m_{\stau} > 500\ {\rm GeV}$
\cite{sev04},  which makes it difficult to test this mechanism at the next
generation of colliders.

Another mechanism is based on the thermal production of gravitinos,
which is determined by the Boltzmann equation,
\begin{equation} 
\frac{dY_{3/2}}{dT} \propto
\frac{\alpha_3(T)}{M_P^2}\frac{m_{\tilde{g}}^2}{m_{3/2}^2}\ .
\end{equation}
This yields the relic gravitino density
\begin{equation}
\Omega_{3/2}h^2\simeq 0.2 
\left(\frac{T_R}{10^{10}\rm{GeV}}\right)
\left(\frac{\rm{100GeV}}{m_{3/2}}\right)
\left(\frac{m_{\tilde{g}}(\mu)}{\rm{1TeV}}\right)^2\ .
\end{equation}
It is quite remarkable that the observed CDM density is obtained for
typical SUSY breaking parameters, $m_{3/2} \sim 100\ {\rm GeV}$,
$m_{\tilde{g}} \sim 1\ {\rm TeV}$ and $T_R \sim \sqrt{m_{3/2}M_P} \sim
10^{10}\ {\rm GeV}$.

In general, the relic density of thermally produced gravitinos depends on
the reheating temperature, and therefore on the initial conditions in the
early universe. There is, however, an interesting case where this dependence
disappears. If the gauge coupling depends on a dilaton field, $g=g(\phi)$,
as in higher dimensional theories, and the dilaton mass is controlled by
supersymmetry breaking, i.e., $m^2_{\phi} = \xi m_{3/2}^2$, 
$\xi = {\cal O}(1)$, then the relic gravitino density becomes essentially
independent of the reheating temperature as well as the gravitino mass
\cite{bhr03}.
  
\begin{figure}
\begin{tabular}{cc}
  \includegraphics[height=.21\textheight]{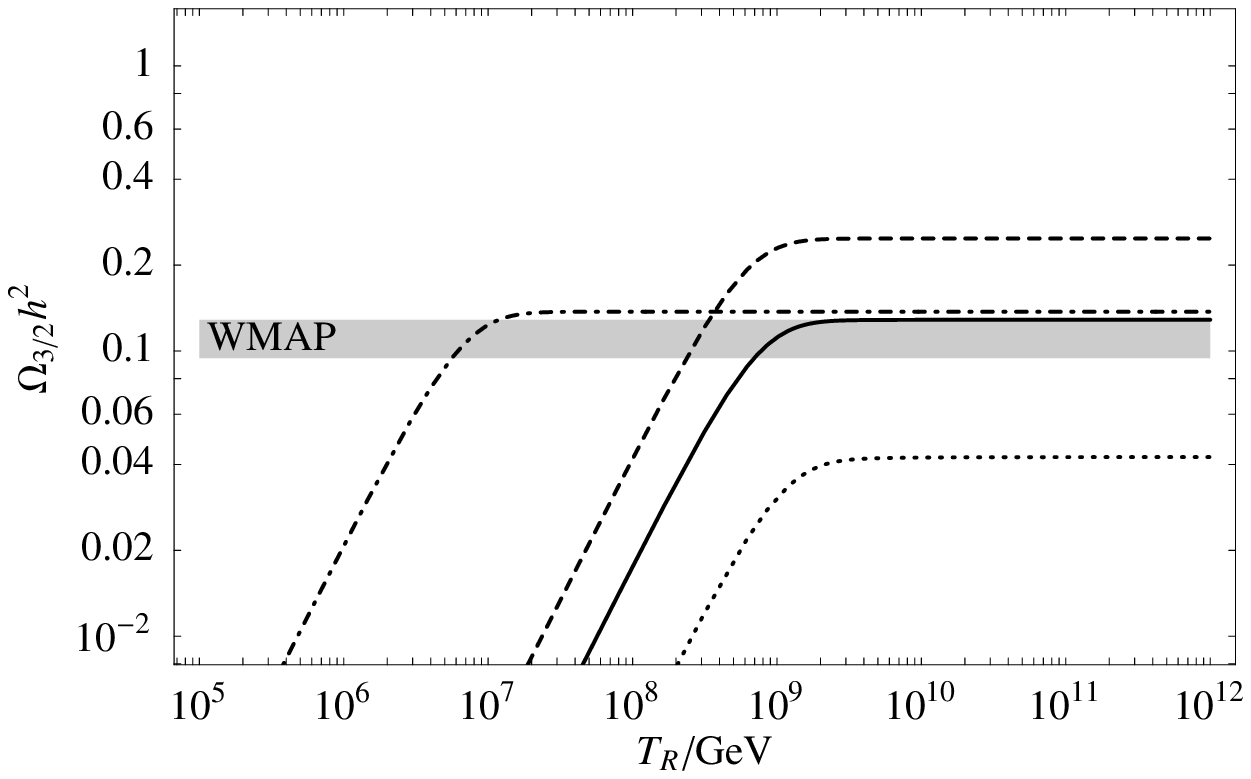} &
  \includegraphics[height=.22\textheight]{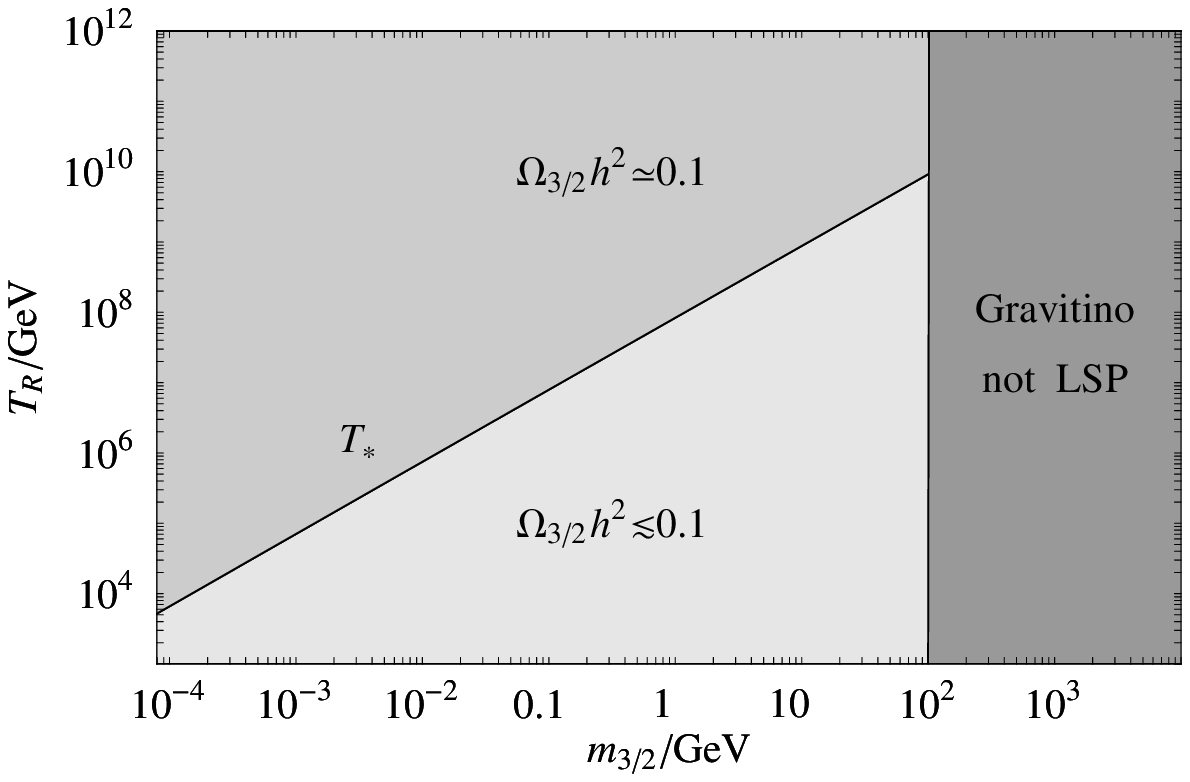}
  \caption{Left panel: Relic gravitino density $\Omega_{3/2}h^2$ as function 
of the reheating temperature $T_R$ for different gravitino and gluino masses: 
$m_{3/2} = 20$~GeV with $m_{\widetilde{g}} = 1.5$~TeV (dashed line), 
$m_{\widetilde{g}} = 1.0$~TeV (full line), 
$m_{\widetilde{g}} = 0.5$~TeV (dotted line), and $m_{3/2} = 200$~MeV with 
$m_{\widetilde{g}} = 1.0$~TeV (dashed-dotted line); $\Omega_{3/2}h^2$ reaches 
a plateau at $T_R \simeq T_* $. 
Right panel: Relic gravitino density for different values of reheating 
temperature and gravitino mass; for $T_R > T_*$, 
$\Omega_{3/2}h^2$ is independent of $T_R$ and $m_{3/2}$. From \cite{bhr03}.}
\end{tabular}
  \label{nllp}
\end{figure}

The gauge coupling multiplies the gauge kinetic term,
\begin{equation}
{\cal L}_{eff} = 
\frac{1}{g^2(\phi)}
 \left(-\frac{1}{4}F^a_{\mu\nu}F^{a\mu\nu}-
         i\lambda^a\sigma^{\mu}(D_{\mu}\bar{\lambda})^a\right) + \dots\ .
\end{equation} 
whose positive expectation value at high temperature drives the effective
gauge coupling to smaller values. This effect sets in at a critical
temperature
\begin{equation}
 T_* \sim \xi^{1/4}
\left({m^2_{3/2} M_{\rm P}\over 2 m_{\tilde{g}}}\right)^{1/2} \;,
\end{equation}
which compensates the increase of the production cross section with
temperature. As a consequence, the gravitino production saturates and becomes
independent of $T$ above $T_*$, yielding
\begin{equation}
\Omega_{3/2}h^2 =  0.1 \times
  \left(\frac{m_{\tilde{g}}(1~{\rm TeV})}{1~{\rm TeV}}\right)^{3/2}
  \xi^{1/4}I_{(\alpha)} F(T_*)\ ,
\end{equation}
with the model dependent parameter $I_{(\alpha)} F(T_*) = 0.5 \ldots 2$. It is
very remarkable that the observed amount of dark matter is obtained for 
$m_{\tilde{g}} \sim 1\ {\rm TeV}$. A shortcoming  of this mechanism is 
a `moduli problem', for which so far no fully satisfactory solution has been 
found.  

The described effect of decreasing gauge couplings at high temperatures leads
to a simple picture for the generation of matter in the universe. Leptogenesis
requires a reheating temperature $T_R$ which is larger than $T_*$
The generated gravitino mass density is then independent
of $T_R$ and $m_{3/2}$. The observed value, 
$\Omega_{3/2} \sim \Omega_{\rm DM}$,
is obtained for $m_{\tilde{g}} \sim 1\ {\rm TeV}$, which will soon be tested
at the LHC. 

\begin{figure}
\begin{tabular}{c}
  \includegraphics[height=.35\textheight]{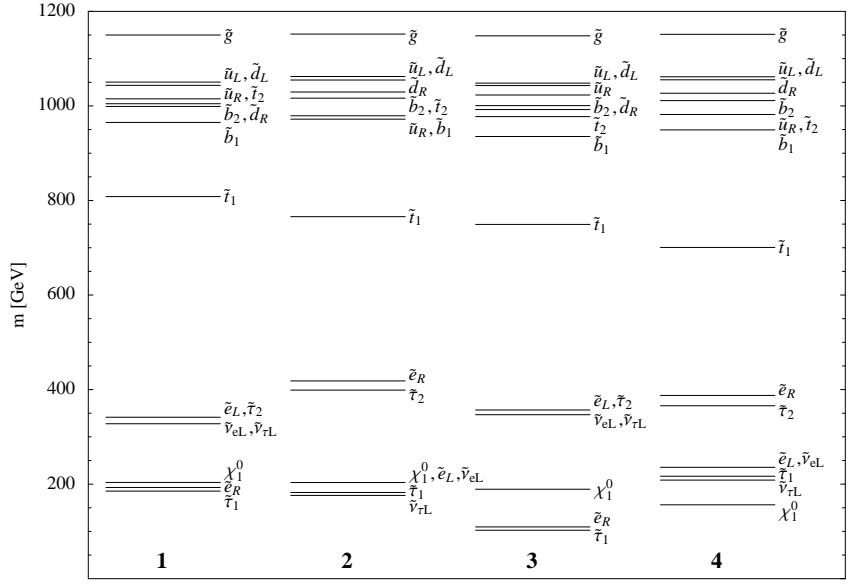} 
  \caption{Four superparticle mass spectra in gaugino mediation with
different NLSP's. From \cite{bks05}.} 
\end{tabular}
  \label{nllp}
\end{figure}

\section{GRAVITINO SIGNATURES AT FUTURE COLLIDERS}

How can one detect gravitino dark matter? To distinguish gravitinos from
WIMPs in cosmological observations is certainly very difficult! It is
therefore encouraging that in the case of a $\tilde{\tau}$-NLSP, 
the next generation of colliders, LHC and ILC, has the potential to discover
the gravitino \cite{bhx04}.

Sufficiently slow, strongly ionizing, long-lived $\tilde{\tau}$'s may be 
stopped. In the $2$-body decay $\tilde{\tau} \rightarrow \tau \gravitino$ 
the gravitino mass can then be kinematically determined from
$m_{3/2}^2 = m_{\tilde{\tau}}^2 + m_\tau^2 -2m_{\tilde{\tau}} E_\tau$,
with the same accuracy as $E_\tau$ and $m_{\tilde{\tau}}$, i.e., a few GeV.
Within supergravity, the measurement of the $\tilde{\tau}$-lifetime then 
yields a determination of the Planck mass:
\begin{equation}
M_{\rm P}^2({\rm supergravity})  = 
 \frac{\left( m_{\tilde{\tau}}^2 - m_{3/2}^2 \right)^4 }{
        48\pi\,m_{3/2}^2\,m_{\tilde{\tau}}^3\, \Gamma_{\tilde{\tau}}}\ .
\end{equation}
Agreement with the `macroscopic' Planck mass,
\begin{equation}
M_{\rm P}^2({\rm gravity}) = (8\pi\ G_{\rm N})^{-1} 
  = (2.436(2)\cdot 10^{18}\ {\rm GeV})^2\ ,
\end{equation}
would be impressive evidence for supergravity! In more refined experiments
it may even be possible to measure the spin of the gravitino. During the past 
two years several groups have studied the feasibility to test supergravity at
LHC and ILC \cite{col04,mar06}.

\begin{figure}
\begin{tabular}{cc}
  \includegraphics[height=.28\textheight]{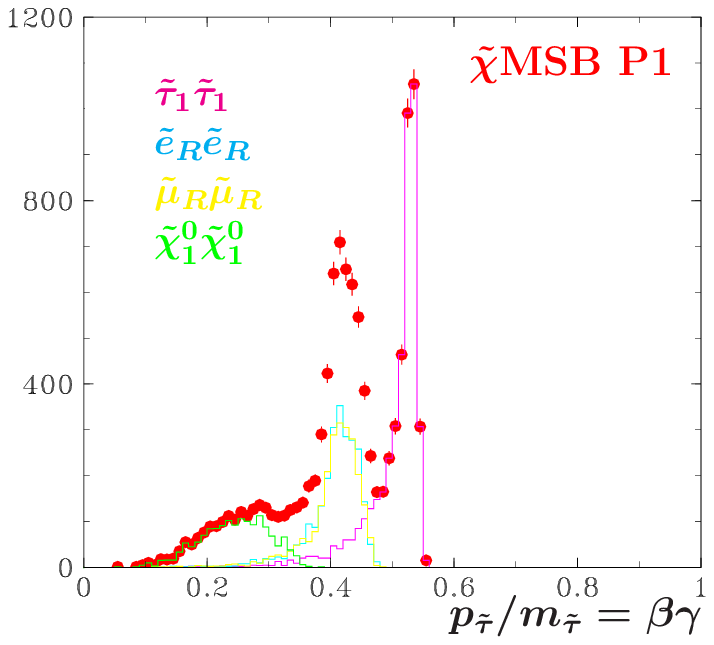} &
  \includegraphics[height=.28\textheight]{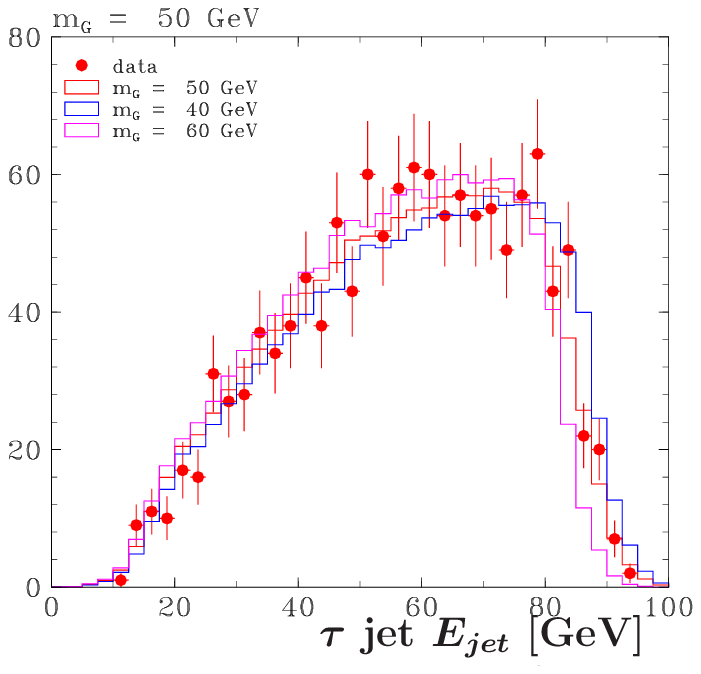}
  \caption{Left panel: Momentum spectrum of $\stau$ leptons. Right panel:
$\tau$-jet energy spectrum in $\stau$ decays. From \cite{um06}.}
\end{tabular}
  \label{led}
\end{figure}
The gravitino is the LSP in models with gaugino mediated supersymmetry 
breaking for a wide range of parameters, with a mass $m_{3/2} > 10\ {\rm GeV}$ 
\cite{bhk05}. The NLSP can be either the lighter $\stau$, the $\snu$ or
a neutralino. Four examples, with different choices of parameters, are shown
in Fig.~6 \cite{bks05}. For the parameter set P1, for instance, one has 
$m_{\tilde{\tau}} = 185.2\ {\rm GeV}$ and $t_{\tilde{\tau}} = 
9.1\times 10^6\ {\rm s}$ for the choice $m_{3/2} = 50\ {\rm GeV}$.
\begin{table}[b]
  \centering
  \renewcommand{\arraystretch}{1.1}
  \begin{tabular}{|c||c|c|c|}
    \hline
      $m_{\tilde{\tau}}$ [GeV] & $t_{\tilde{\tau}}$ [s] & 
       $m_{3/2}(\Gamma_{\tilde{\tau}}) [GeV]$ & $m_{3/2}(E_\tau)$ [GeV] \\ 
    \hline 
    $185.2 \pm 0.1$ & $(9.1 \pm 0.2)\times 10^6$ & $50 \pm 0.6$ & $50 \pm 3$\\ 
    \hline 
  \end{tabular}
\caption{Results of a Monte Carlo study for the determination of $m_{\stau}$,
$t_{\stau}$ and $m_{3/2}$ at the ILC. From \cite{mar06}.} 
\end{table}
Recently, a detailed Monte Carlo study has been carried out for the ILC
\cite{mar06}. Different mass patterns of superparticles have been studied,
including the spectrum P1 of gaugino mediation. Several production processes 
generate $\stau$'s with the momentum spectrum shown in Fig.~7. For an 
integrated luminosity ${\cal L} \sim 200\ {\rm fb^{-1}}$ at $\sqrt{s} =420\ 
{\rm GeV}$ one expects about $4\times 10^3$ slow ($\beta < 0.3$) $\stau$'s
to be stopped
in the hadron calorimeter (HCAL). The $\stau$ mass and lifetime can be
measured rather accurately. From the endpoint of the $\tau$-jet energy the
gravitino mass $m_{3/2}(E_{\tau})$ can be determined with an accuracy of a
few GeV. This implies that also the Planck mass can be determined 
`microscopically' with an error of about 10\% !


In summary,
thermal leptogenesis requires a large reheating temperature in the early
universe. In supersymmetric theories this leads to strong constraints on
the allowed superparticle mass spectrum. The standard supergravity scenario
with an unstable gravitino with mass ${\cal O}(1\ {\rm TeV})$ is imcompatible
with thermal leptogenesis. A consistent picture is obtained with a 
gravitino LSP which can be the dominant component of dark matter. In the
case of a $\stau$ NLSP, the next generation of colliders, LHC and ILC,
has the potential to discover the gravitino and to establish 
spontaneously broken local
supersymmetry as a hidden symmetry of nature.\\ 

\noindent
I am grateful to my collaborators on `leptogenesis and gravitino dark matter', 
M.~Bolz, A.~Brandenburg, L.~Covi,
P.~Di~Bari, K.~Hamaguchi, M.~Ibe, J.~Kersten, R.~D.~Peccei, 
M.~Pl\"umacher, M.~Ratz, K.~Schmidt-Hoberg and T.~Yanagida.


\bibliographystyle{aipprocl} 

\end{document}